\documentclass[conference]{IEEEtran}
\IEEEoverridecommandlockouts
\usepackage{cite}
\usepackage{amsmath,amssymb,amsfonts}
\usepackage{algorithmic}
\usepackage{graphicx}
\usepackage{textcomp}
\usepackage{blindtext}
\usepackage{float} 
\usepackage{subfigure}
\usepackage{xcolor}
\usepackage{wrapfig}
\usepackage{makecell}

\def\BibTeX{{\rm B\kern-.05em{\sc i\kern-.025em b}\kern-.08em
		T\kern-.1667em\lower.7ex\hbox{E}\kern-.125emX}}
\makeatletter

\begin{document}
	\title{Effect of Retransmissions on the Performance of C-V2X Communication for 5G}
	\IEEEpeerreviewmaketitle
	\author{\IEEEauthorblockN{Donglin Wang, Raja R.Sattiraju, Anjie Qiu, Sanket Partani and Hans D. Schotten}
		\IEEEauthorblockA{\textit{University of Kaiserslautern} \\
			Kaiserslautern, Germany \\
			$\{$dwang,sattiraju,qiu,partani,schotten$\}$@eit.uni-kl.de}
	}
	
	\maketitle
\begin{abstract}
In recent years, the next generation of wireless communication (5G) plays a significant role in both industry and academy societies. Cellular Vehicle-to-Everything (C-V2X) communication technology has been one of the prominent services for 5G. For C-V2X transmission mode, there is a newly defined communication channel (sidelink) that can support direct C-V2X communication. Direct C-V2X communication is a technology that allows vehicles to communicate and share safety-related information with other traffic participates directly without going through the cellular network.
The C-V2X data packet will be delivered to all traffic Users (UE) in the proximity of the Transmitter (Tx). Some UEs might not successfully receive the data packets during one transmission but the sidelink Tx is not able to check whether the Receivers (Rxs) get the information or not due to the lack of feedback channel. For enabling the strict requirements in terms of reliability and latency for C-V2X communication, we propose and evaluate one retransmission scheme and retransmission with different traffic speed scheme. These schemes try to improve the reliability of the safety-related data by one blind retransmission without requiring feedback. Although this retransmission scheme is essential to C-V2X communication, the scheme has a limitation in the performance aspect because of its redundant retransmission. Since radio resources for C-V2X communication are limited, we have to detect the effect of retransmission on the performance of the communication system. To the end, the simulator for evaluating the proposed schemes for the C-V2X communication has been implemented, and the performances of the different communication schemes are shown through the Packet Reception Ratio (PRR).  
\end{abstract}


\section{Introduction}

In modern society, many issues like big traffic congestion, energy consumption, and air pollution merge with the development of the transportation system [1]. To address these issues, the 3rd Generation Partnership Project (3GPP) Release 14 specification published Vehicle-to-Everything (V2X) specification based on Long-Term Evolution (LTE) as the underlying technology which provides communication services in vehicular scenarios. There is the Cellular-based V2X communication which it's generally referred to as C-V2X to differentiate itself from the 802.11p based V2X technology [2]. C-V2X is the introduced technology for optimizing transportation and connected vehicles. It promises to transform safety-related and efficiency-related information on highways and within cities both by connecting individual vehicles and by enabling the development of a Cooperative-Intelligent Transportation System (C-ITS), which can reduce congestion and pollution, enhance travel efficiency, and avoid traffic collisions [3].
The key concept of C-V2X communication lies in sidelink and lots of researches proposed direct C-V2X communication [4][5][6]. 
And direct C-V2X communication is using broadcast operating mode. The broadcast could use multi-hop transmissions to enhance coverage [4], but a single-hop transmission is suggested in recent studies and it’s applied to this work. In the direct C-V2X communication, data packets are transmitted directly from the Tx to the Rxs in the proximity of the Tx without going through the network infrastructures. So sidelink enables some essential vehicular services which have a high low-latency requirement, such as the transmission of Cooperative Awareness Messages (CAMs) and Decentralized Environmental Notification Messages (DENMs) [7].
The characteristics of the communication via sidelink are basically derived from the communication via LTE uplink [8].
For example, UEs transmit information through sideling based on the framework of the uplink. From another aspect, the sidelink has an individual characteristic which is no feedback channel for reporting whether the transmitted information from the Tx has been received or not concerning sidelink. Consequently, the UEs are not able to have Hybrid Automatic Retransmission Request (HARQ) ACK/NACK information for the sidelink. Instead, the sidelink allows the UEs to perform HARQ retransmission blindly.

There are quite lots of works about direct transmissions with retransmission technologies [5][9].
[5] proposed index-coded retransmissions for enhancing sidelink channel efficiency of V2X communications and [9] evaluated the effect of retransmissions on the performance of the IEEE 802.11p protocol for Dedicated Short Range Communication (DSRC). 

In this work, we implement one HARQ blind retransmission on the system-level of the direct C-V2X communication for enabling the strict requirements in terms of reliability for C-V2X communication. Moreover, we have an SNR-BLER mapping table from link-level simulator [10]. The scheme tries to improve the reliability of the safety-related data by one blind retransmission without requiring feedback. This retransmission scheme is essential to C-V2X communication. The scheme has a limitation in the performance aspect due to the redundant retransmission. Since there is limited radio resources for C-V2X communication, one blind retransmission with varying traffic speeds is applied to check the system performance. 

\begin{figure}[htbp]
	\centering
	\includegraphics[width=\linewidth]{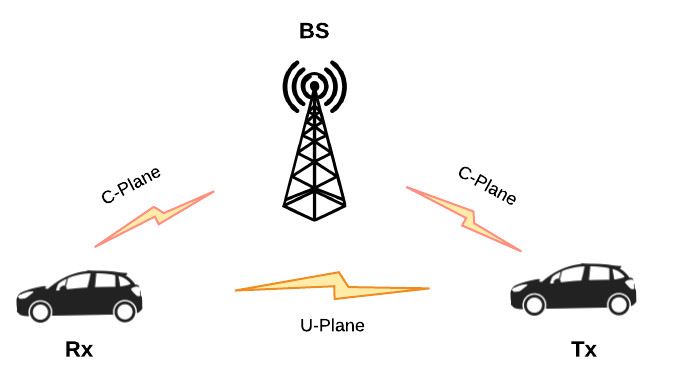}
	\caption{Direct C-V2X communication with network assistance}
	\label{fig}
\end{figure}

\section{System mode of direct C-V2X communication  }
Direct C-V2X communication through sidelink is a mode of communication whereby a UE can directly communicate with other UEs in its proximity over the PC5 air interface. This communication is a point-to-multipoint communication where several Rxs try to receive the same data packets transmitted from a transmitting UE. In the present section, we discuss system-level simulation results obtained for network-assisted direct C-V2X transmission in a highway traffic scenario, as illustrated in fig. 1. All UEs are connected to BSs of the same Radio Access Network (RAN). Scheduling and resource control information are transmitted from BSs to the UEs via the Vehicle-to-network (V2N) Control-plane (C-plane) of the radio interface. The transmitting Tx directly transmits its Vehicle-to-Vehicle (V2V) or Vehicle-to-Infrastructure (V2I) sidelink data packets to the surrounding Rxs with its communication range, thereby achieving low latency. 
The RAN can provide network control for direct C-V2X communication. In 3GPP-defined C-V2X, there are two alternative sidelink transmission modes:
	\begin{itemize}
	\item Sidelink transmission mode 3
	\end{itemize}
	In this mode, the resource allocation for each sidelink transmission is scheduled by a BS. This transmission mode is only available when the vehicles are under cellular coverage. To assist the resource allocation procedure at the BS, UE context information (e.g., traffic pattern information) can be reported to BSs.
	
	\begin{itemize}
	\item Sidelink transmission mode 4
	\end{itemize}
	In this mode, a Tx in C-V2X communication can autonomously select a radio resource from a resource pool which is either configured by network or pre-configured in the user device for its direct C-V2X communication over PC5 interface. In contrast to mode 3, transmission mode 4 can operate without cellular coverage. 
	
	\begin{figure*}[htbp]
		\centering
		\subfigure[Without retransmission]{
			\includegraphics[width=0.45\textwidth,height=0.7\columnwidth]{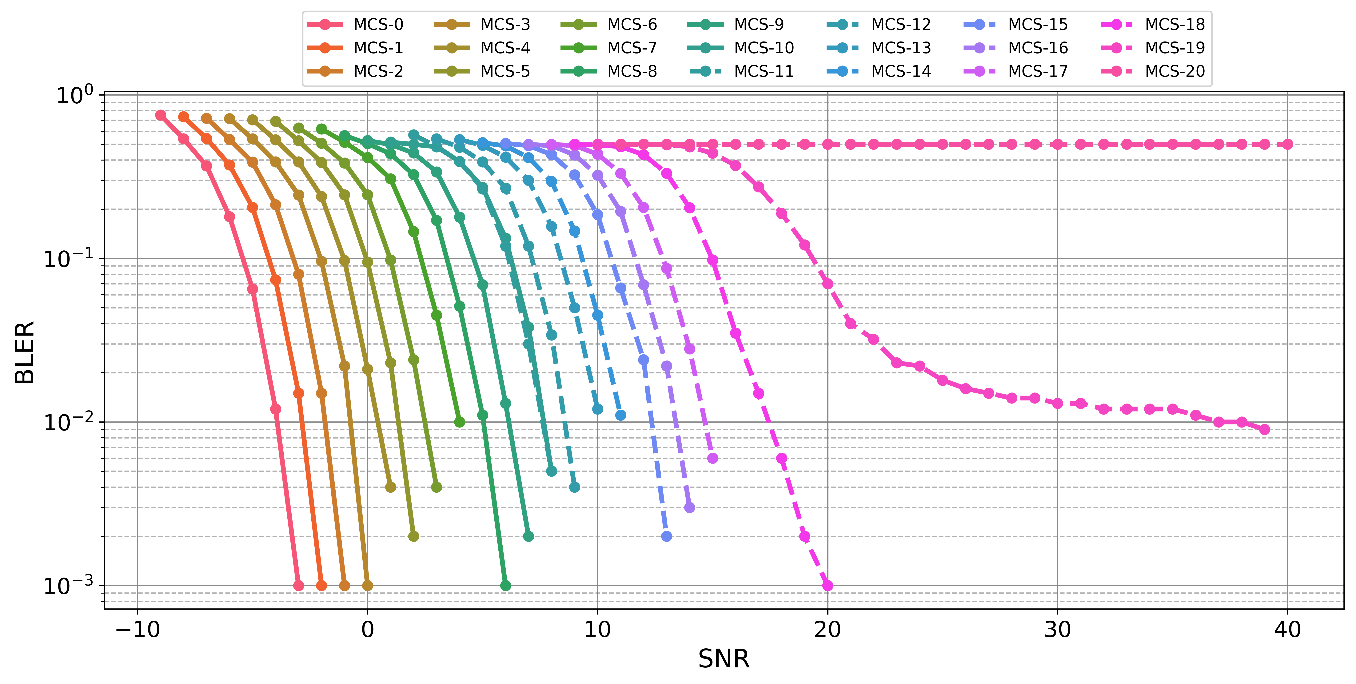}}
		\label{fig.sub.1}
		\subfigure[With retransmission]{
			\includegraphics[width=0.45\textwidth,height=0.7\columnwidth]{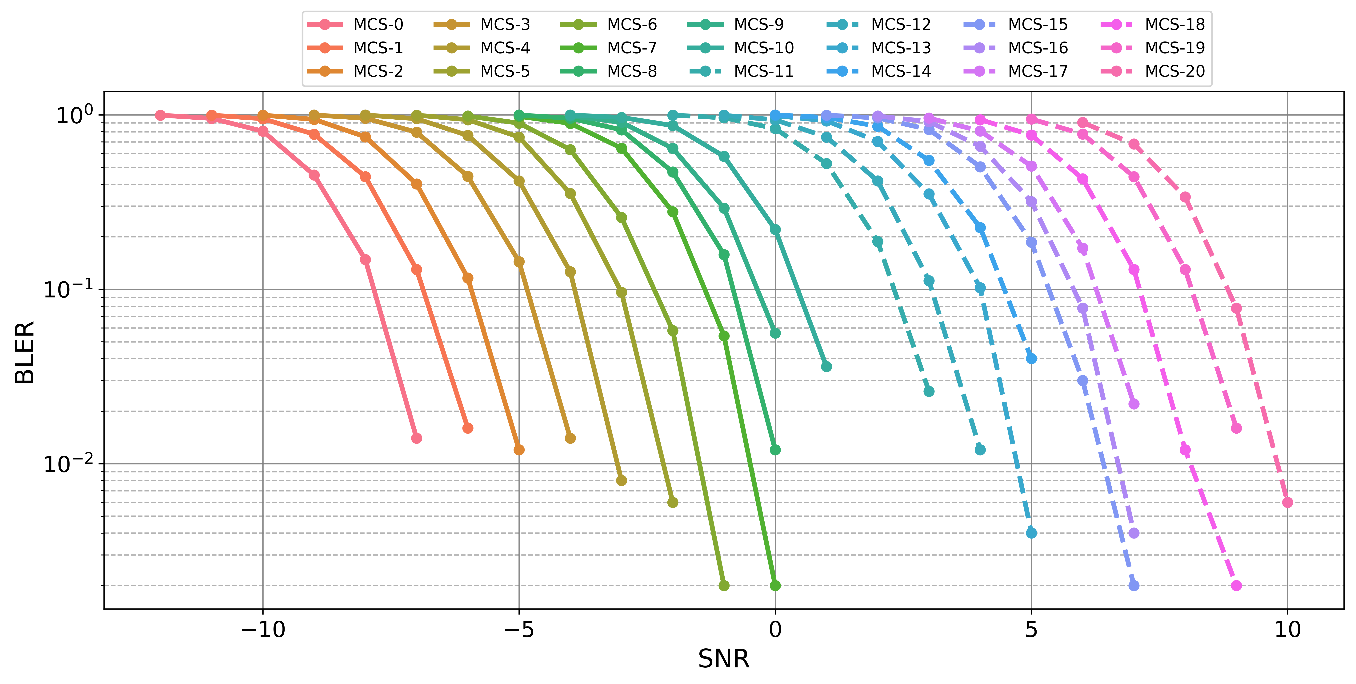}}
		\label{fig.sub.2}
		\caption{Direct C-V2X Performance for EVA Channel at 100 km/h}
		\label{fig}
	\end{figure*}
	
In the system-level simulation analysis described below, transmission mode 3 is utilized for the direct C-V2X communication through sidelink which means all UEs are under the coverage of a cellular network that controls the resource allocation of sidelink transmissions.

Moreover, the system-level simulation mainly reflects the Media Access Control (MAC) layer functionalities such as resource allocation, user scheduling, and adaptive Modulation and Coding Scheme (MCS), rather than the physical layer processing. The performance of the physical link is taken into account by a Link-to-System (L2S) interface derived from link-level simulations [10]. Moreover, to precisely reflect the characteristics of a radio link (e.g., frequency fading), an L2S mapping needs to be accurately formulated. Mutual information-based L2S is one of the commonly used methods which has been considered as preferable and applicable. Physical-layer procedures have to be abstracted by accurate but also low-complexity models. 
For our system-level simulations, we used the following L2S mapping tables: 
\begin{enumerate}
	\item The mapping table that considers all MCS(0-20) for the ITU-Extended Vehicular A (EVA) channels fading channel at 100 km/h speeds. No blind retransmission is considered. This mapping is denoted as L2S-1 respective curves are given in fig. 2(a) [10]. 
	\item The mapping table that considers the same channel and speeds as L2S-2 but with one blind retransmission. This mapping is denoted L2S-2 and respective curves can be found in fig. 2(b) [10].
\end{enumerate}
In our system-level simulation, resource allocation, mobility management, admission control, interference management, HARQ, and scheduling are modelled. The Key Performance Indicators (KPIs) are Signal-to-Noise-plus–Interference-Ratio (SINR), MCS, BLER, and the PRR [11]. We consider the traffic consists of CAM messages, where each message is contained in a single packet, which in turn is contained in a single transport block. For this reason, the PRR is simply the inverse of the BLER.
The PRR is calculated from all Rxs within the intended communication range of the Tx. For each transmitted CAM message, the PRR can be calculated as X/Y, where Y is the total number of UEs located in the communication range from the UE transmitting the message, and X is the number of UEs in that range that successfully receives the message [1]. The average PRR is calculated as:
\begin{equation}
PRR=\frac {X_1 + X_2 + \cdots + X_n}	{Y_1 + Y_2 + \cdots + Y_n} \label{eq}
\end{equation}
where the index represents the message for which the reception is evaluated and n is the total number of messages in the simulation. When a UE is transmitting, i.e. is in the role of a Tx UE, in practice, it cannot receive at the same time. Since multiple packets may be transmitted by other Tx UEs at the same time, on different frequency resources, a UE may miss multiple packets while it is itself transmitting. The missed packets due to this restriction are, however, not reflected in the PRR in the system simulator, there are more system configurations in [11].

\section{Sidelink system-level simulation assumptions}
In this section, we highlight in detail the simulation assumptions.
\subsection{Environment Model}
Two BSs are deployed with an Inter-Site-Distance (ISD) of 1732 m alongside the 3464-meter highway scenario with 6 lanes to provide control for the UEs of the C-V2X communication. The 2 BSs are centered along the highway, i.e. the BSs are placed at +/-1732m/2 from the horizontal center of the highway. 
Due to the limited number of cells and due to our methodology in the first scenarios where we only consider the UEs that are in-between the 2 BSs, the interference in the simulated scenario will be substantially smaller than in practice and the results are therefore optimistic. In one simulation scenario, the highway length is extended to 6928-meter, with 4 BSs deployed, so that the edge area in which interference is reduced represents a rather small fraction of the total area.
UEs are deployed with a fixed Inter-Vehicle-Distance (IVD) on each lane at the beginning of the simulation. During the simulation, no UEs are entering or leaving. We assume the desired communication range of the UEs is 400 m [2]. Results are presented in the range of $IVD$ = 5 m to 100 m independent of vehicle speed. Note that in realistic traffic scenarios IVD in meters is typically equal or larger than half the vehicle speed measured in kilometers per hour, e.g. $IVD$ $\geq$ 50 m at speed of 100 km/h. 
The number of UEs on the highway scenario has been calculated as:
\begin{equation}
UE_{H} =\frac{L_H}{IVD} \times Lane \label{eq}
\end{equation}
Where $L_H$ is the length of the highway scenario, $Lane$ is the number of highway lanes, and $IVD$ is the inter vehicle distance. 
\subsection{traffic model}
The number of the UEs controlled by each BS is calculated as:
\begin{equation}
UE_{BS} =\frac{ISD}{IVD} \times Lane \label{eq}
\end{equation}
Where $ISD$ is the $ISD$ and $IVD$ and Lane are defined as in the equation before.
We assume each UE transmits a packet of 256 bytes 10 times per second on average in most scenarios. The packet transmission rate is a variable parameter in some of the results we will show further below.
The data volume (in bits per second) is derived as follows: 
\begin{equation}
C =ps \times 8 \times UE_{BS} \times P \label{eq}
\end{equation}
where $ps$ is the packet size of 256 bytes, $P = 10$ is the average number of packet transmissions per second, and $UE_{BS}$ is obtained from the equation before. 
\subsection{resource allocation and scheduling}
Direct C-V2X communication in transmission mode 3 uses resources in so-called resource pools. The resource pools are configured by the BS to its controlled UEs. Rx UEs monitor the Physical Sidelink Control Channel (PSCCH) in the configured resource pool for transmissions from Tx UEs. The BS precisely specifies the time-frequency resources to be used by each of its controlled Tx UEs. The Rx UEs are not informed about this by the BS. They will get the information from the monitored PSCCHs. Typically, and in our simulation scenarios, the resource pools configured by all BSs are the same. In this way, it is ensured, that the Rx UEs can receive transmissions from Tx UEs controlled by any BS. In particular, in our simulation, there is one resource pool in each cell and that covers the entire time and frequency domain.

Each BS ensures that it allocates each available resource to only a single one of its controlled transmitting UEs, so that for an Rx UE, interference can only emerge from Tx UEs controlled by a different BS. 

This is a conservative resource usage strategy, because in principle it may make sense for a BS to allocate a resource to multiple Tx UEs if they are sufficiently far apart. There are, however, no standardized procedures on how the BS could know for which combination of Tx UEs would be beneficial. Therefore, resource reuse among UEs controlled by one BS is not considered here. Furthermore, with an intended communication range of 400 m and an ISD of 1732 m, there is little room from a geographic perspective to have 2 Tx UEs transmit on the same resource without overlap between their communication ranges or overlap with that of a Tx UE controlled by a neighboring BS [11].

In the simulations, the time-frequency resource grid of LTE is not modelled explicitly. After the UE deployment, the simulation runs for 1000 iterations. In each iteration, for each of the BSs, one of the UEs that are controlled by the BS is selected randomly with equal probability for transmission. This represents a fully loaded system. The Rx UEs within the communication range of the selected Tx UE is then evaluated for the packet reception, regardless of by which BS an Rx UE is controlled. If the system is overloaded, some UEs cannot be supported by BS, and are therefore dropped. If the system is not fully loaded according to the Inter-Vehicle-Distance(IVD) and message rate [11], then in a real system there would be iterations (subframes) where no UE controlled by one BS does transmit, thereby the total interference generated in such iterations would be reduced. In our simulation, however, there is one Tx UE per BS in each iteration, thereby overestimating the interference. For the practically relevant cases of a message rate of 10 Hz, however, there is, in fact, one Tx UE per BS in each iteration even for the highest considered IVD of 100 m.
For the case of no-retransmissions, we calculate the SINR ($\gamma_1$) for each Rx-UE in the communication range for a selected Tx and look up the L2S-1 mapping for the corresponding BLER, using constant-value extrapolation for SINR values not covered by the link simulation range. A random number X is then generated from a uniform distribution [0, 1]. If $X < BLER$, then the packet is marked as received.
\subsection{resource allocation for one blind retransmission of direct C-V2X communication }
\begin{figure}[htbp]
	\centering
	\includegraphics[width=\linewidth]{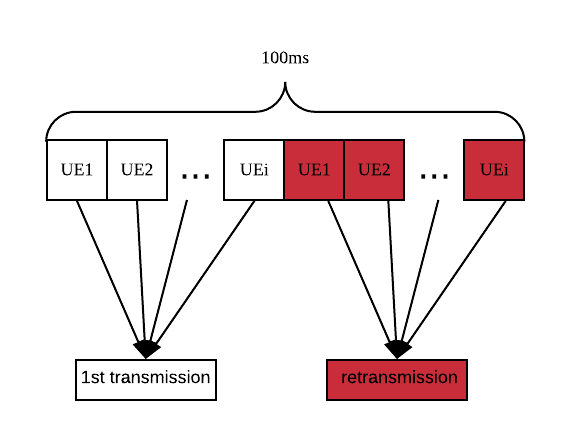}
	\caption{resource allocation}
	\label{fig}
\end{figure}

To improve the performance of direct C-V2X communication, one blind retransmission is implemented in the simulation. 
As fig. 3 shown, in this example, $U_1$, $U_2$, $\cdots$, $U_i$ are served by one BS in 100 ms (duration time of one period of 10 Hz). we assume that first transmission and retransmission occupy 50 ms resource respectively and the number of $i$ UEs is supported by BS according to the IVD and message rate. 
 
As mentioned before, in case of a retransmission, $\gamma_1$ is calculated in the same way as outlined previously. Then, another interfering Tx is chosen randomly and a new SINR value ($\gamma_2$) is calculated for each Rx-UE in the communication range for the same selected Tx. Both these SINR values are averaged to get the final SINR, i.e., $\gamma_{final}=(\gamma_1+\gamma_2)/2 $ (in linear scale). The corresponding BLER is looked up from the L2S-2 table. Similar to the previous process, a random number $X$ is then generated from a uniform distribution [0, 1]. If $X > BLER$, then the packet is marked as received.

Taking the average SINR over the first transmission and the retransmission is necessary because the L2S-2 table considers SINR values that are averaged over different SINR realizations. In the link simulation, each realization represents a different fast fading channel realization, whereas in the light of a system simulation with different interferers the averaging also has to consider the differing SINR corresponding to interferer realizations.

\subsection{channel model}
In this work, isotropic antennas are installed on the top of each vehicle at a height of 1.5 m. A 1×2 antennas configuration is exploited for direct C-V2X communication. Also, each Tx has assumed a constant Equivalent Isotropically Radiated Power (EIRP) of 23 dBm. According to 3GPP specifications, power control is applied where the transmit power depends on the used transmission bandwidth and the distance between the UE and the controlling BS. The first part is not relevant in our simulation model as the full transmission bandwidth is always assumed to be used. The second part leads to a transmit power variation that is uncorrelated to the sidelink pathloss. This causes an increase in the SIR variance and accordingly lower SINR at the lower tail of the SINR distribution, so this tends to lead to somewhat worse performance in reality than in the simulation. 
The central carrier frequency is 5.9 GHz with a transmission bandwidth of 10 MHz [2]. The WINNER II models [2] are applied as propagation models for calculating the pathloss. The UEs in the system simulation are static and do not move. A time-varying channel is however taken into account by the link-to-system model.
\subsection{modulation and coding scheme}
Since there are different MCSs supported in C-V2X, the network needs to configure the appropriate MCSs for each transmission. An appropriate MCS should meet the data volume requirement 
\begin{equation}
SE \geq \frac{C} {BW} \label{eq}
\end{equation}

\begin{figure*}[htbp]
	\centering
	\subfigure[100 km$/$h without transmission]{
		\includegraphics[width=0.45\textwidth,height=0.7\columnwidth]{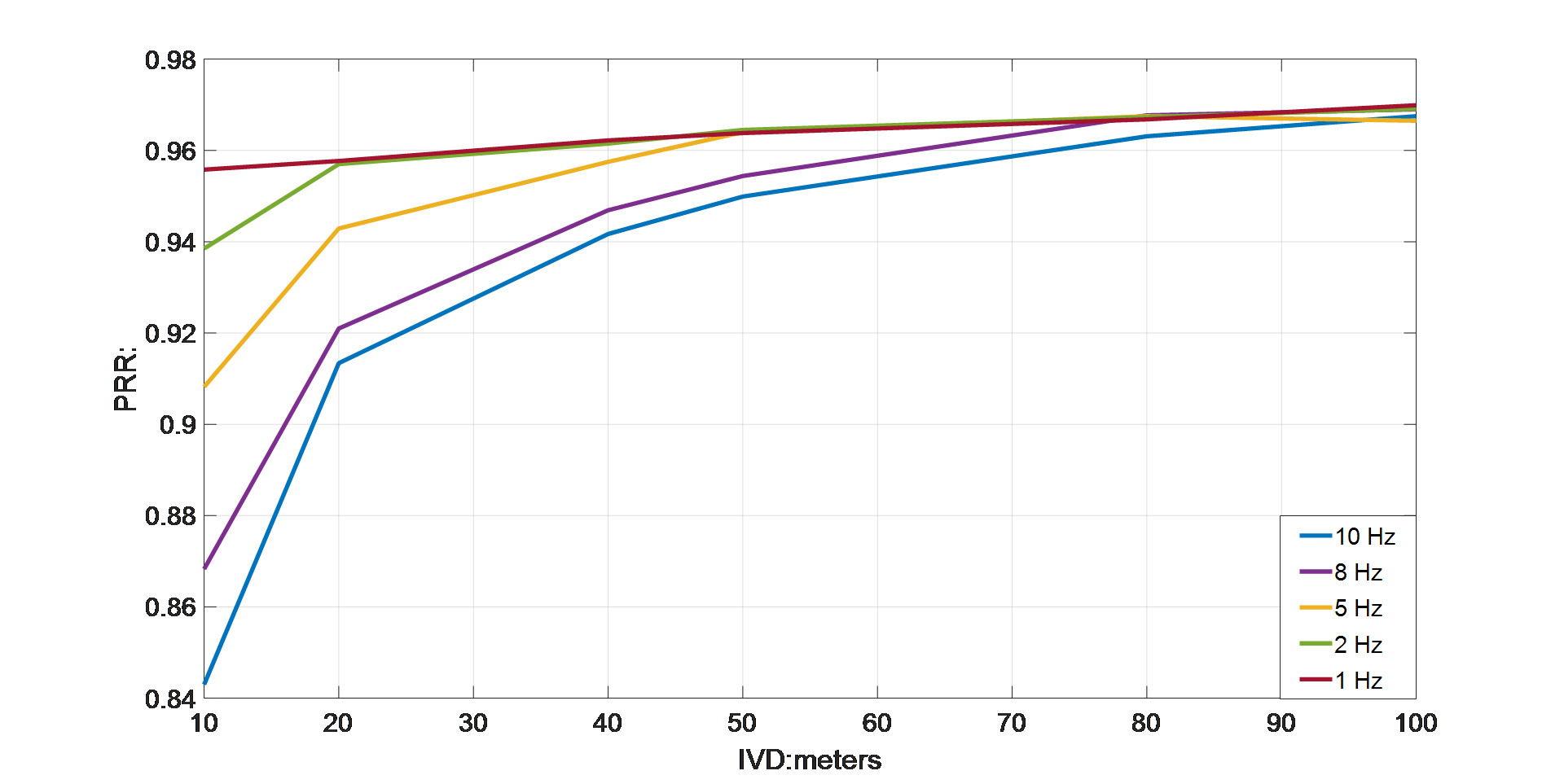}}
	\label{fig.sub.1}
	\subfigure[100 km$/$h with retransmission]{
		\includegraphics[width=0.45\textwidth,height=0.7\columnwidth]{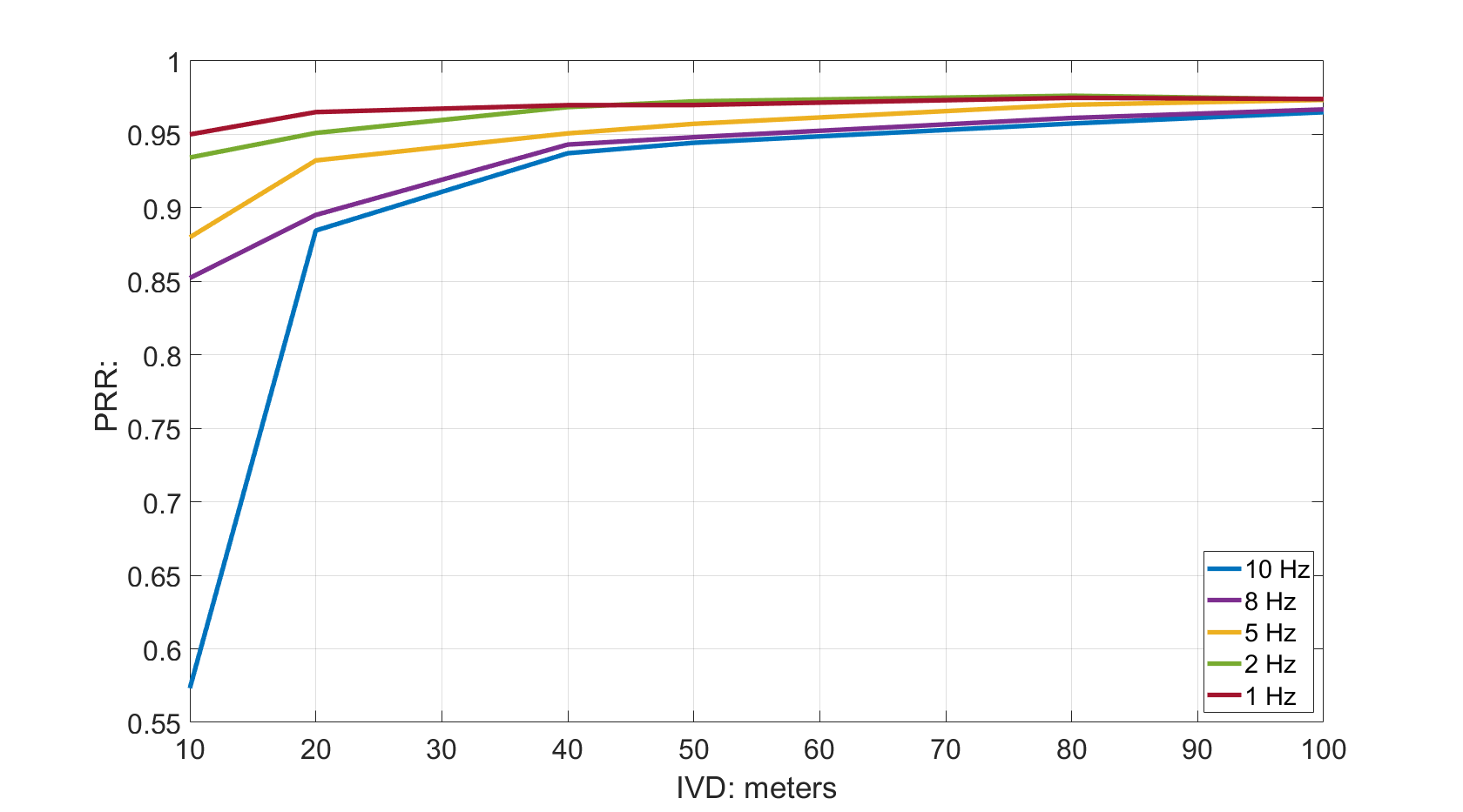}}
	\label{fig.sub.2}
	\subfigure[10 Hz without retransmission ]{
		\includegraphics[width=0.45\textwidth,height=0.7\columnwidth]{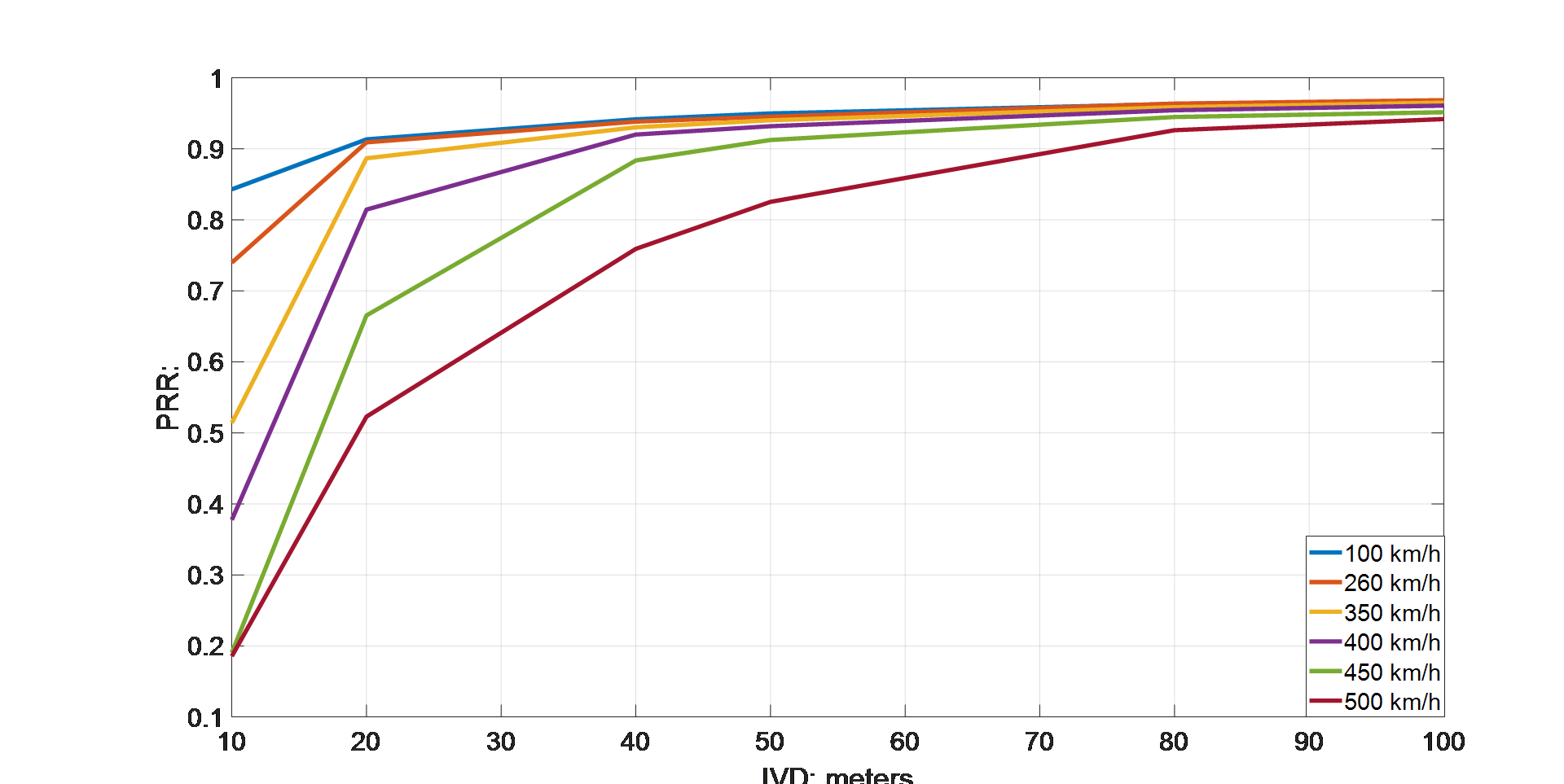}}
	\label{fig.sub.3}
	\subfigure[10 Hz with retransmission]{
		\includegraphics[width=0.45\textwidth,height=0.7\columnwidth]{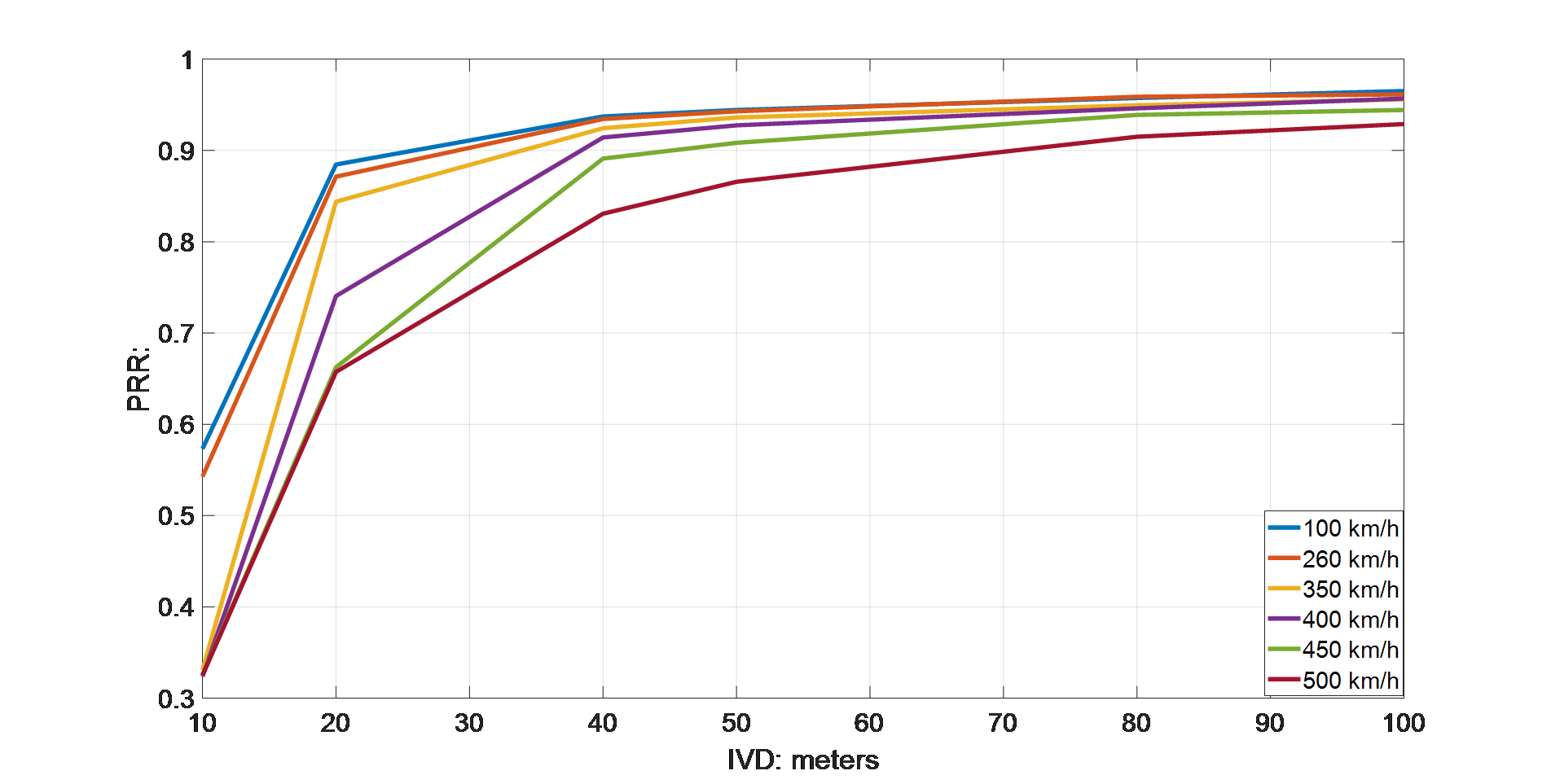}}
	\label{fig.sub.4}
	\caption{PRR vs. IVD with and without retransmission for EVA channel}
	\label{fig}
\end{figure*}

where $SE$, $C$, and $BW$ represent the spectral efficiency of an MCS, the data volume as defined before, and the allocated bandwidth, respectively. Additionally, since an MCS with a higher spectral efficiency is less robust, the MCS which has the lowest spectral efficiency while fulfilling the condition shown in Eq. (5) should be applied. For $IVD$ $\leq$ 5, however, even MCS 20 with the highest spectral efficiency does not provide compliance with condition (5), and consequently the system is overloaded. This will be addressed further below.
Table.I shows the relations of UE deployment density, data volume and selected MCS index for the case of no retransmission.
\begin{table}[htbp]
\caption{UE deployment, data volume and  MCS Parameters for scenarios without retransmission}
\begin{center}
	\begin{tabular}{|p{0.5cm}<{\centering}|p{1.5cm}<{\centering}|p{1.5cm}<{\centering}|p{1.5cm}<{\centering}|p{1.5cm}<{\centering}|}
		\hline 
		IVD (m) &	$\#$UEs on highway&	$\#$UEs on BS coverage&	Data volume [Mbps]&	MCS index (L2S-1)\\
		\hline
		3&		6928&	3464&	70.9427&	20\\  
		\hline
		5&		4156&	2078&	42.5574&	20\\  
		\hline
		10&		2078&	1039&	21.2787&	14\\  
		\hline
		20&		1039&	519&	10.6291&	7\\  
		\hline
		40&		519&	259&	5.3043&		4\\  
		\hline
		50&		415&	207&	4.2394&		3\\  
		\hline
		80&		259&	129&	2.6419&		1\\  
		\hline
		100&	207&	103&	2.1094&		0\\  
		\hline
	\end{tabular}
\end{center}
\end{table}

For the case of retransmissions, the data volume is effectively doubled, and accordingly the required spectral efficiency of the MCS is doubled. This is shown in Table I. For $IVD$ $\leq$ 10 m even the MCS 20 with the highest spectral efficiency does not provide compliance with condition (2), and consequently, the system is overloaded. 

\begin{table}[htbp]
\caption{UE deployment, data volume and MCS Parameters for retransmission}
\begin{center}
	\begin{tabular}{|p{0.5cm}<{\centering}|p{1.5cm}<{\centering}|p{1.5cm}<{\centering}|p{1.5cm}<{\centering}|p{1.5cm}<{\centering}|}
		\hline 
			IVD (m) &	$\#$UEs on highway&	$\#$UEs on BS coverage&	Data volume [Mbps]&	MCS index (L2S-2)\\
		\hline
		3&		6928&	3464&	141.8854&	20\\  
		\hline
		5&		4156&	2078&	85.1148&	20\\  
		\hline
		10&		2078&	1039&	42.5574&	20\\  
		\hline
		20&		1039&	519&	21.2582&	14\\  
		\hline
		40&		519&	259&	10.6086&	7\\  
		\hline
		50&		415&	207&	8.4788&		6\\  
		\hline
		80&		259&	129&	5.2838&		4\\  
		\hline
		100&	207&	103&	4.2188&		3\\  
		\hline
	\end{tabular}
\end{center}
\end{table}
 
In overload scenarios, the BS is unable to support all the UEs and hence drops some of them. The PRR calculation in this case also considers the dropped UEs. Let $PRR_{max}$ denote the maximal PRR achievable for an overloaded scenario, which is given as the ratio of a total number of supported UEs to the total number of UEs within the coverage area of BS:
\begin{equation}
PRR_{max} \frac{UE_{supported}} {UE_{BS}} \label{eq}
\end{equation}
During simulation runtime, the $PRR$ is calculated only considering the $UE_{supported}$. Let us denote it as $PRR_{runtime}$. $PRR_{runtime}$ is defined as a percentage of UEs that successfully receive a packet from the tagged Tx among the Rxs within the transmission range of the Tx in the running time as shown in Eq.1. The final effective PRR that is shown in the subsequent figures in this chapter is then calculated by multiplying the runtime PRR with the maximum PRR as follows 
\begin{equation}
PRR = \frac{PRR_{max}} {PRR_{runtime}} \label{eq}
\end{equation}

Table III shows the values of $PRR_{max}$ for different IVD values with and without the use of retransmissions. It can be seen that overloading only happens for $IVD$ $<$ 10 $\:$ m for the case of no retransmission and $IVD$ $<$ 20 $\:$ m for the case with retransmission.
\section{results analysis}
The system simulation has been carried out for EVA channel for various speeds. Also, the simulation is repeated considering one blind retransmission. 
In Fig. 4(a), the PRRs of the different transmission frequencies for the sidelink C-V2X communication are plotted. It is easy to find that the PRR of sidelink C-V2X communication increases with raising IVDs from 10 m to 100 m, due to the accompanying decrease in MCS index. The PRR value increases from 84.30$\%$ to 96.75$\%$  when the IVD is increased from 10 m to 100 m with 10 Hz transmission rate. 
In fig. 4(b), The PRR value increases from 57.32$\%$ to 96.51$\%$  when the IVD is increased from 10 m to 100 m with 10 Hz transmission rate with one blind retransmission. 
The PRR for different message rates appears to converge about 0.96 with increasing IVD. We provide some contemplations on the expected impact of IVD on the PRR: (1) The wanted received signal power on average is not affected, because it is always evaluating over the same communication range of 400 m. However, within the same lane, as the IVD is constant, the distance to the furthest Rx within the communication range does depend on the IVD, but not in a monotonous way. Since each lane has a random offset from the leftmost UE to the left edge of the simulation, the cross-lane wanted received power is further randomized and therefore a noticeable effect of the IVD is not expected; (2) Within the same lane, the minimum distance between an Rx and interfering Tx is equal to the IVD, so with increasing IVD the PRR of the worst Rxs should increase. For cross-lane, again due to random offset, the minimum distance can be as small as the lane separation for any IVD, it is just so that for smaller IVD the smaller distances become more probable, but each Tx UE on a small distance has a lower activity ratio because there is always one active Tx per cell, so the smaller the IVD the more UEs you have the less often one UE is active. So for larger IVD, having an interfering Tx UE at a small distance is less probable, but if there is one then it is interfering more often; (3) The MCS decreases with increasing IVD because it is determined only according to the traffic load. This has the largest effect on the PRR. From Table I without ReTx the MCS = 0 is reached for IVD = 100 m, for 10 Hz, so that means the MCS cannot further reduce with further increasing IVD, therefore the PRR is expected to saturate, as the SINR discussion above also does not reveal a clear improvement trend. For lower message rates that MCS 0 is simply reached already for lower IVDs.
Fig.4.(c) and 4.(d) show the PRR performance for different speeds for a message frequency of 10 Hz with and without retransmission. The following points can be noted: 

1. For $IVD$ $<$ 10 $\:$m, the retransmissions case results in an overloaded scenario as seen by the low PRR. This is also visible in Table II where the highest MCS is selected for $IVD$ = 10 $\:$m. Due to this, the maximum PRR is limited to the total number of supported UE’s divided by the total number of UE’s present in the network.

2. At high speeds (and consequently high Doppler frequencies), retransmissions have a larger benefit compared to lower speeds. This can be seen in the form of a steeper PRR curve where the retransmission gain is 10 - 15$\%$ over the case with no retransmissions for every IVD until 70 m. Above 70 m, due to the lower MCS used, the performance converges.

3. Some aspects of the link and system simulation scenario are expected to lead to lower gains from retransmissions than expected. The first is that always full bandwidth 10 MHz transmissions are assumed, which implies a high degree of frequency diversity to that the time diversity introduced by retransmissions is less relevant. The second is that the system simulation assumes a fully loaded system where there is interference present always in the retransmission, whereas for a different resource allocation scheme this would not be the case.

\section{conclusion}
On the system level, this report has investigated C-V2X "mode 3", where the resource allocation for the direct V2V communication is performed by BSs. For a message rate of 10 per second, the PRR is above the 90$\%$ target for all IVD above 40 m and UE speeds up to 400 km/h when the MCS is optimally chosen. For 500 km/h this is only the case for IVD above 80 m, however, given the high speed, the breaking distance does also require such high IVD. 

With a single retransmission, the load is naturally doubled, and accordingly less robust MCS is used. There are hardly any simulation scenarios where retransmissions can lift the PRR above the target of 90$\%$. However, some aspects of the link and system simulation scenario are expected to contribute to that the retransmissions do not harvest the full diversity gain potential. At high speeds retransmissions do increase the PRR by 10 - 15$\%$, however, still only at an unacceptably low level. 

The baseline system simulation results have been obtained for a system consisting of only 2 BSs and with some simulation-friendly restrictions on UE deployment. However, for the most relevant scenarios the simulations have been repeated with larger system size and realistic UE deployment and BS association and the PRR declined by at most 2$\%$-points. The conclusions seen from the system level analysis on LTE-V2X sidelink “mode-3” is that it suffers from performance issues in addition to a number of added complex requirements and business considerations.
\section{Acknowledgement}
A part of this work has been supported by Federal Ministry of Transport and Digital Infrastructure (BMVI) in the framework of the project ConVeX with funding number 16AVF1019. The authors would like to acknowledge the contributions of their colleagues, although the authors alone are responsible for the content of the paper which does not necessarily represent the project.

\end{document}